\documentclass[journal=jacsat,manuscript=article]{achemso}

\usepackage[version=3]{mhchem} 
\usepackage{color}
\usepackage{rotating}
\usepackage{lscape}
\usepackage{url}
\usepackage{xr}
\usepackage{cleveref}
\usepackage{longtable}
\usepackage{threeparttable}
\usepackage{threeparttablex}
\usepackage{array}
\usepackage{float}

\usepackage{xr}
\usepackage{xc}

\newcolumntype{C}[1]{>{\centering\arraybackslash}p{#1}}
\newcolumntype{R}[1]{>{\raggedleft\arraybackslash}p{#1}}
\newcolumntype{L}[1]{>{\raggedright\arraybackslash}p{#1}}

\makeatletter
\newcommand*{\addFileDependency}[1]{
  \typeout{(#1)}
  \@addtofilelist{#1}
  \IfFileExists{#1}{}{\typeout{No file #1.}}
}
\makeatother
 
\newcommand*{\myexternaldocument}[1]{%
    \externaldocument{#1}%
    \addFileDependency{#1.tex}%
    \addFileDependency{#1.aux}%
}

\myexternaldocument{ccsdt_frequencies_SI}

\setkeys{acs}{articletitle = true}

\author{Luke W. Bertels}
\affiliation{Department of Chemistry, University of California, Berkeley, California 94720, USA.}
\author{Joonho Lee}
\affiliation{Department of Chemistry, University of California, Berkeley, California 94720, USA.}
\alsoaffiliation{Department of Chemistry, Columbia University, New York, New York 10027, USA.}
\author{Martin Head-Gordon}
\affiliation{Department of Chemistry, University of California, Berkeley, California 94720, USA.}
\alsoaffiliation{Chemical Sciences Division, Lawrence Berkeley National Laboratory, Berkeley, California 94720, USA.}
\email{mhg@cchem.berkeley.edu}

\title{Polishing the Gold Standard: The Role of Orbital Choice in CCSD(T) Vibrational Frequency Prediction
}

\abbreviations{CCSD, CCSD(T), CCSD(T), RHF, ROHF, UHF, MP2, OOMP2, $\kappa$OOMP2, MR, RMSD, MSD, MIN, MAX}

\begin{document}

\begin{tocentry}
\includegraphics[height=3.5cm]{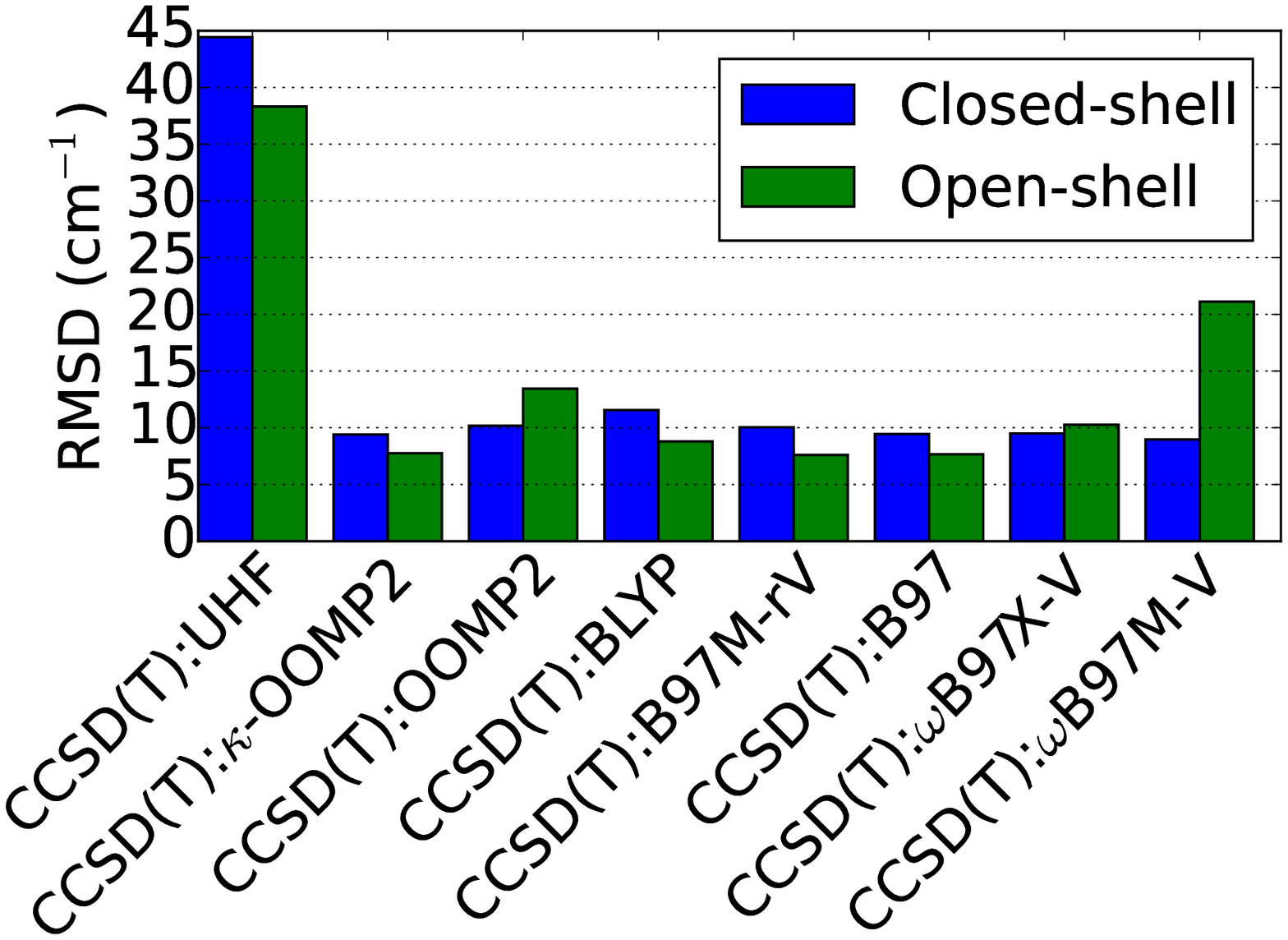}




\end{tocentry}

\begin{abstract}
 While CCSD(T) with spin-restricted Hartree-Fock (RHF) orbitals has long been lauded for its ability to accurately describe closed-shell interactions, the performance of CCSD(T) on open-shell species is much more erratic, especially when using a spin-unrestricted HF (UHF) reference. Previous studies have shown improved treatment of open-shell systems when a non-HF set of molecular orbitals, like Brueckner or Kohn-Sham density functional theory (DFT) orbitals, is used as a reference. Inspired by the success of regularized orbital-optimized second-order M{\o}ller-Plesset perturbation theory ($\kappa$-OOMP2) orbitals as reference orbitals for MP3, we investigate the use of $\kappa$-OOMP2 orbitals and various DFT orbitals as reference orbitals for CCSD(T) calculations of the corrected ground-state harmonic vibrational frequencies of a set of 36 closed-shell (29 neutrals, 6 cations, 1 anion) and 59 open-shell  diatomic species (38 neutrals, 15 cations, 6 anions). The aug-cc-pwCVTZ basis set is used for all calculations. The use of $\kappa$-OOMP2 orbitals in this context alleviates difficult cases observed for both UHF orbitals and OOMP2 orbitals. Removing 2 multireference systems and 12 systems with ambiguous experimental data leaves a pruned data set. Overall performance on the pruned data set highlights  CCSD(T) with a B97 orbital reference (CCSD(T):B97), CCSD(T) with a $\kappa$-OOMP2 orbital reference (CCSD(T):$\kappa$-OOMP2), and CCSD(T) with a B97M-rV orbital reference (CCSD(T):B97M-rV) with RMSDs of 8.48 cm$^{-1}$ and 8.50 cm$^{-1}$, and 8.75 cm$^{-1}$ respectively, outperforming CCSD(T):UHF by nearly a factor of 5. Moreover, the performance on the closed- and open-shell subsets show these methods are able to treat open-shell and closed-shell systems with comparable accuracy and robustness. The use of $\kappa$-OOMP2 orbitals has also proven useful in diagnosing multireference character that can hinder the reliability of CCSD(T). 
 
\end{abstract}

\section{Introduction}
Coupled cluster theory with single, double, and perturbative triple excitations [CCSD(T)]\cite{raghavachari1989fifth} with spin-restricted Hartree-Fock (RHF) orbitals is considered the ``gold standard" by many quantum chemists for its ability to routinely provide results approaching chemical accuracy for energies and properties of closed-shell species at a reasonable computational cost \cite{thomas1993balance,helgaker1997prediction,bak2001accurate}. With the ability to achieve sub-kcal mol$^{-1}$ errors at a computational cost of $\mathcal{O}(N^{7})$ and a memory cost of $\mathcal{O}(N^{4})$, where $N$ is the size of the basis, CCSD(T) strikes an advantageous accuracy-to-cost balance between coupled cluster theory with single and double excitations (CCSD)\cite{purvis1982full} ($\mathcal{O}(N^{6})$ computational cost, $\mathcal{O}(N^{4})$ memory cost) and explicit treatment of the triple excitations (CCSDT)\cite{noga1987full} ($\mathcal{O}(N^{8})$ computational cost, $\mathcal{O}(N^{6})$ memory cost). For open-shell species, however, the performance of CCSD(T) (especially on top of spin-unrestricted HF (UHF) orbitals) is less clear. A study of bond lengths and frequencies of 33 small radical species by Byrd et. al\cite{byrd2001theoretical} reported that CCSD(T) with a UHF reference (CCSD(T):UHF) shows little to no statistical improvement over CCSD\cite{purvis1982full} for geometries and frequencies. Beran et al.\cite{beran2003approaching}, in a study of harmonic vibrational frequencies of diatomic radicals, reported poor behaviour of CCSD(T):UHF for several species in the set including, \ce{CO+} and \ce{NO}. Subsequent studies of these problem systems by Szalay et. al\cite{szalay2004triplet} investigated the source of this mixed performance. Stanton and Gauss\cite{stanton2003discussion} reviewed several potential factors leading to this discrepancy in describing open-shell species, including multireference (MR) character, spin-contamination, symmetry breaking, and instabilities and near-instabilities in the reference wavefunction. The use of restricted open-shell HF (ROHF) orbitals has been found to somewhat improve the performance on vibrational frequencies over UHF, though the former are prone to errors due to spatial symmetry breaking\cite{stanton2003discussion,tentscher2012geometries}. While in the limit of full configuration interaction the energy and properties are invariant to the choice of reference orbitals, any truncated approximate method will incur some level of orbital dependence.     

Several strategies have been proposed as alternative references to UHF for CC calculations. The use of Brueckner orbitals in CC theory (BCC), which by definition have no singles contribution to the coupled cluster wavefunction, attempts to incorporate the most important electron correlations at the level of the reference\cite{brueckner1954nuclear,nesbet1958brueckner,dykstra1977examination,handy1989size,hampel1992comparison}. BCC approaches have been known to preserve wavefunction symmetries as well, yielding more accurate properties\cite{stanton1992choice,barnes1994symmetry,xie1996oxywater,crawford1997c}. In a similar vein, the molecular orbitals (MOs) can be directly optimized in the presence of CCD correlation energies as in the orbital-optimized coupled cluster doubles (OD) and orbital optimized coupled-cluster doubles and perturbative triples [OD(T)] approaches\cite{scuseria1987optimization,krylov1998size,van2000quadratic}. The use of OD(T) by Beran et al.\cite{beran2003approaching} was shown to significantly suppress errors in the computed harmonic frequencies of \ce{CN}, \ce{CO+}, and \ce{NO} compared to CCSD(T):UHF. Brueckner and/or optimized coupled cluster orbitals are costly to obtain; BCCD(T) and OD(T) typically are far more computationally expensive than CCSD(T) as orbital optimization is often more challenging than varying the singles amplitude. Given this steep computational cost, methods to approximate Brueckner or otherwise optimized orbitals at a lower cost are highly desirable. 

One such approximate approach to incorporate correlation into the reference orbitals for a correlated calculation is the use of Kohn-Sham density functional theory (DFT) orbitals as a reference. The use of BLYP\cite{becke1988density,lee1988development} orbitals by Beran et al.\cite{beran2003approaching} was shown to substantially improve the computed vibrational frequencies of radical diatomic species over UHF orbitals. More recently, Fang et al.\cite{fang2016use} and Fang et al.\cite{fang2017prediction} have demonstrated the efficacy of CCSD(T) with DFT orbitals in the prediction of thermodynamic properties of \ce{UCl6} and several diatomic transition metal compounds, respectively. Aside from their use in CC calculations, DFT orbital references have been used successfully for excited state configuration interaction\cite{bour2001configuration}, as guiding functions in quantum Monte Carlo \cite{needs2002quantum}, and for second-order perturbation theory in the context of double-hybrid DFT\cite{grimme2006semiempirical2}. In addition to their inexpensive computational cost ($\mathcal{O}(N^{3}$)), DFT orbitals offer improved stability against symmetry breaking compared to HF orbitals\cite{sherrill1999performance}. A connection between DFT orbitals and Brueckner orbitals has been proposed by several researchers as well\cite{scuseria1995connections,hesselmann2002first,lindgren2002brueckner}.

Orbital-optimized second-order M{\o}ller-Plesset perturbation theory (OOMP2) and its variants offer another way to approximate higher-order orbital optimized-methods at a cost of $\mathcal{O}(N^{5})$ per iteration\cite{lochan2007orbital,neese2009assessment,lee2018regularized}. Orbital optimization at the MP2 level, in addition to improving energetics, is often seen to reduce spin-contamination in the optimized reference\cite{lochan2007orbital,neese2009assessment,soydas2015assessment}. 
Recently Haggag et al.\cite{haggag2020elusive} utilized OOMP2 reference orbitals for CC calculations on the triplet state of permanganate to combat spin contamination seen at the UHF level. Despite these benefits, OOMP2 exhibits three unsatisfying characteristics that limit its application: divergence in the cases of small orbital energy gaps\cite{stuck2013regularized}, the loss of Coulson-Fischer points\cite{coulson1949notes}, and ``artificial" symmetry restoration\cite{lee2019distinguishing, leethesis}.  The correlation energy functional for MP2 is
\begin{equation}
    E_{\textrm{MP2}} = -\frac{1}{4}\sum_{ijab}\frac{\left|\langle ij||ab\rangle\right|^{2}}{\Delta_{ij}^{ab}}
\end{equation}
where $\Delta_{ij}^{ab} = \epsilon_{a} + \epsilon_{b} - \epsilon_{i} - \epsilon_{j}$ is the non-negative orbital energy denominator. This energy is seen to diverge in cases where the denominator becomes small, as can occur when stretching bonds. This behavior leads to poor performance of OOMP2 when predicting harmonic vibrational frequencies\cite{stuck2013regularized}. Secondly, OOMP2 often fails to continuously transition from restricted to unrestricted solutions even when the unrestricted solution is lower in energy\cite{sharada2015wavefunction}. Thirdly, OOMP2 has been shown in some cases to ``artificially" restore spin-symmetry to systems where the spin symmetry breaking is an ``essential" feature of the system due to MR character of the system\cite{lee2019distinguishing,leethesis}. 

In order to address the problematic aspects of OOMP2, two of us developed $\kappa$-OOMP2, a regularized variant of OOMP2\cite{lee2018regularized}. The $\kappa$-OOMP2 energy functional is given by 
\begin{equation}
    E_{\kappa\textrm{-OOMP2}}(\kappa) = -\frac{1}{4}\sum_{ijab}\frac{\left|\langle ij||ab\rangle\right|^{2}}{\Delta_{ij}^{ab}}\left(1-e^{-\kappa\Delta_{ij}^{ab}}\right)^{2},
\end{equation}
where $\kappa$ is a regularization parameter that damps contributions to the correlation energy when the orbital energy denominator becomes small. Regularization parameter values $\kappa \leq 1.5$ $E_{h}^{-1}$ were shown to restore Coulson-Fischer points for a series of bond-breaking curves\cite{lee2018regularized}. Training of the regularization parameter on the W4-11 thermochemistry data set\cite{karton2011w4} led to an optimal $\kappa$ value of 1.45 $E_{h}^{-1}$\cite{lee2018regularized}. With this parameter value, $\kappa$-OOMP2 was able to outperform OOMP2 on the TA13 data set\cite{tentscher2013binding} of radical--closed-shell interaction energies\cite{lee2018regularized}. Further application to symmetry breaking in fullerenes revealed the ability of $\kappa$-OOMP2 to distinguish between essential and artificial symmetry breaking\cite{lee2019distinguishing,leethesis}. 

We recently developed a scaled variant of third-order MP theory (MP3) that utilizes $\kappa$-OOMP2 orbitals as a reference which we will denote as  MP2.8:$\kappa$-OOMP2\cite{bertels2019third}. MP2.8:$\kappa$-OOMP2 and its unscaled version, MP3:$\kappa$-OOMP2, outperformed CCSD on five of the seven data sets investigated at the cost of a single $\mathcal{O}(N^{6})$ iteration. The use of $\kappa$-OOMP2 orbitals strongly improves upon the performance of MP3 with UHF orbitals as well, especially in cases of spin-symmetry breaking.  

Inspired by MP2.8:$\kappa$-OOMP2\cite{bertels2019third}, the work of Beran et al.\cite{beran2003approaching}, and the success of $\kappa$-OOMP2 in treating radical species in the TA13 set\cite{lee2018regularized} and in producing minimally spin-contaminated references for biradicaloid systems\cite{lee2019two}, in this work we explore the use of $\kappa$-OOMP2  orbitals as a reference for CCSD(T) computation of vibrational frequencies. Errors are calculated with respect to experimental values and compared against CCSD(T) with UHF orbitals, OOMP2 orbitals, and several flavors of DFT orbitals.This use of $\kappa$-OOMP2 orbitals as a reference for CCSD(T) was previously explored in the computation of spin-gaps in an iron porphyrin complex and showed an improvement over conventional CCSD(T) \cite{lee2020utilizing}.

\section{Computational Methods}

We consider eight methods as generators of MOs for use as references: UHF, two OOMP2 methods (OOMP2\cite{lochan2007orbital,neese2009assessment,bozkaya2011quadratically} and $\kappa$-OOMP2\cite{lee2018regularized}), and five density functionals (BLYP\cite{becke1988density,lee1988development}, B97M-rV\cite{mardirossian2017use}, B97\cite{becke1997density}, $\omega$B97X-V\cite{mardirossian2014omegab97x}, and $\omega$B97M-V\cite{mardirossian2016wb97m-v}). A regularization parameter value of $\kappa =1.45 E_{h}^{-1}$ was chosen for $\kappa$-OOMP2, as suggested by Lee and Head-Gordon\cite{lee2018regularized}. Both the OOMP2 and $\kappa$-OOMP2 calculations were carried out using the resolution-of-the-identity (RI) approximation\cite{feyereisen1993use,bernholdt1996large}. The functionals B97M-rV, B97, $\omega$B97X-V, and $\omega$B97M-V were chosen on the basis of their performance in a recent benchmark of over 200 density functionals in which they were found to be the best performing meta-GGA, global hybrid GGA, range-separated hybrid GGA, and range-separated hybrid meta-GGA functionals, respectively\cite{mardirossian2017thirty}. The DFT calculations were performed using an ultra-fine integration grid of 99 radial points and 590 angular points per atom. Wavefunction stability analysis\cite{seeger1977self,sharada2015wavefunction} was performed on the UHF and DFT solutions to ensure that all orbitals used properly minimize their corresponding SCF energies. All unstable solutions (saddle points) were displaced and reoptimized to local minima. OOMP2 and $\kappa$-OOMP2 calculations were performed starting from a locally stable UHF solution. Both the reference calculations and CCSD(T) calculations were performed using unrestricted wavefunctions. 

All calculations were performed with the aug-cc-pwCVTZ basis\cite{dunning1989gaussian,kendall1992electron,peterson2002accurate,woon1993gaussian,prascher2011gaussian} set to capture the effects of core correlation. A core-valence basis set was utilized to account for the role of core-valence correlations in molecular property calculations. The use of the weighted, triple-$\zeta$ variant is justified by the faster convergence of properties to the complete basis set limit seen with cc-pwCV$n$Z over cc-pCV$n$Z\cite{peterson2002accurate}. Augmented functions were chosen to better treat the anions in the data set. The corresponding auxiliary basis set was utilized for the OOMP2 and $\kappa$-OOMP2 calculations (with cc-pwCVQZ-RI utilized for \ce{Li}, \ce{Be}, \ce{Na}, and \ce{Mg})\cite{weigend2002efficient,hattig2005optimization,hill2010correlation}. Atoms \ce{H}-\ce{F} have all electrons correlated and atoms \ce{Na}-\ce{Cl} utilize a frozen \ce{He} core. All electronic structure calculations were performed using the Q-Chem package of electronic structure programs\cite{shao2015advances}.
For a given reference, the corrected harmonic vibrational frequency was determined by fitting a quartic polynomial to seven equally-spaced points (0.005 {\AA}  between adjacent points) distributed near the minimum of the potential and applying Dunham analysis to the fitting coefficients to account for the effects of rotation\cite{dunham1932energy}. Equilibrium bond lengths from the fitting procedure are reported in Tab. \ref{tab:bonds_ccsdt_cs} and \ref{tab:bonds_ccsdt_os}.   

\section{Experimental Data Selection}

For simplicity and due to the lack of analytic first derivatives of the CCSD(T) energy with respect to nuclear displacements for non-HF references, we restricted the systems of study in this work to diatomic species for which the ground state potential energy surface can be determined via fitting to single point calculations. Beginning from all diatomics of row 2 and row 3 species (and hydrogen) for which Huber and Herzberg\cite{huber2018constants} report ground state frequencies, we include all species for which we were able to compute a smooth potential energy surface about the corresponding equilibrium bond length for each method (excluded species are listed in Tab. \ref{tab:rejects}). The overall data set contains 36 closed-shell species (29 neutrals, 6 cations, 1 anion) and 59 open-shell species (38 neutrals, 15 cations, 6 anions; 46 doublets, 13 triplets). Among other species in this data set, we include all 12 species from Beran et al.\cite{beran2003approaching} as well as several isovalent analogues of these species containing row 3 elements. Other notable inclusions are \ce{B2} ($X^{3}\Sigma_{g}^{-}$) and \ce{C2} ($X^{1}\Sigma_{g}^{+}$), both known to exhibit MR behavior in their ground states\cite{watts1992coupled,karton2011w4}, and \ce{F2} ($X^{1}\Sigma_{g}^{+}$), a biradicaloid diamagnetic system know that is unbound at the UHF level of theory\cite{purwanto2008eliminating}. Where available, experimental frequencies were updated with data from Irikura\cite{irikura2007experimental}. Frequencies for \ce{OH-}, \ce{F2+}, and \ce{SO+} were updated with more recent experimental data from Hotop et al.\cite{hotop1974high}, Cormack et al.\cite{cormack1996high}, and Milkman et al.\cite{milkman1988high}, respectively.  

\section{Results and Discussion}

\section {Vibrational frequencies}
\begin{figure}
\centering
\caption{Box plots (overall, left, and enhanced, right) of the errors in corrected vibrational frequencies (in cm$^{-1}$) are presented. Red lines mark the median errors, boxes bound the central 50\% of the data, whiskers enclose all data points within 1.5 times the inter-quartile range of the box edges, and points denote outlying data. }
\label{fig:boxplot_all}
\includegraphics[width=\linewidth]{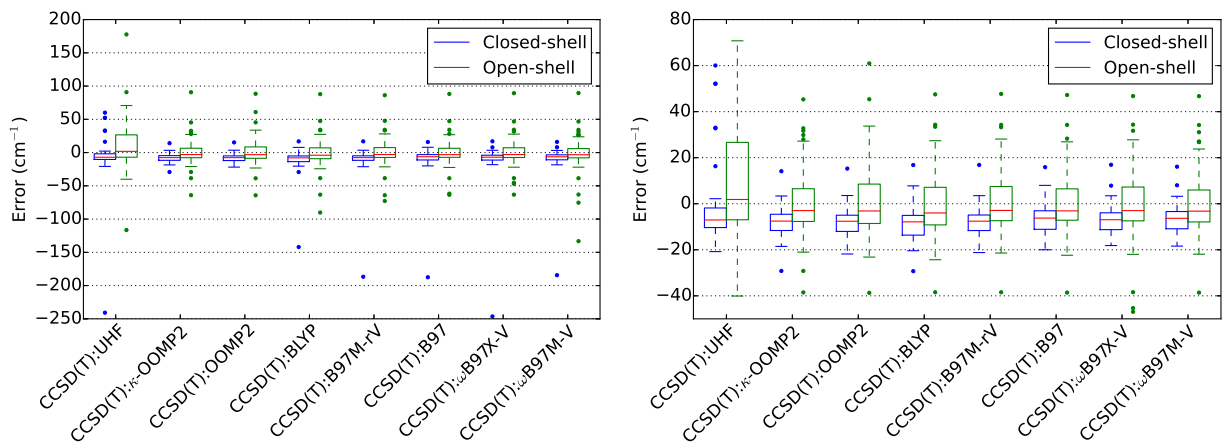}
\end{figure}

Fig. \ref{fig:boxplot_all} presents the errors in the corrected vibrational frequencies on the closed- and open-shell subsets for each method as box plots. In these plots, the boxed region represents data from the first to the third quartile of the distribution, a red line marks the median of the data, whiskers enclose all data within 1.5 times the inter-quartile distance of the upper and lower box edges, and points mark data lying outside of these regions. Tab. \ref{tab:all} presents the root mean square deviations (RMSD), mean signed deviations (MSD), most negative deviations (MIN), and most positive deviations (MAX), all in cm$^{-1}$, of the corrected vibrational frequencies from the experimental frequencies for the overall data set. CCSD(T):$\kappa$-OOMP2 is seen to give the best overall performance in terms of RMSDs with a value of 17.66 cm$^{-1}$, reducing the RMSD for CCSD(T):UHF by more than a factor of two. CCSD(T):OOMP2 is seen to perform slightly worse but still improves on the CCSD(T):UHF RMSD by a factor of 2. The performance of the DFT orbital approaches is hindered by the presence of \ce{C2}, representing the MIN value for all functionals tested. The MAX value, corresponding to \ce{PH+}, is shared among all non-HF methods tested. On average the frequencies are slightly blue-shifted for CCSD(T):UHF and slightly red-shifted for all other methods.

\begin{table}
\caption{Root mean square deviations, mean signed deviations, most negative deviations, and most positive deviations (all in cm$^{-1}$) for the predicted corrected vibrational frequencies of all species are summarized for the CCSD(T) methods utilizing different molecular orbital references.} \label{tab:all}
\tiny
\begin{tabular}{L{1.4cm}R{1.4cm}R{1.4cm}R{1.4cm}R{1.4cm}R{1.4cm}R{1.4cm}R{1.4cm}R{1.4cm}}
\hline
     & \multicolumn{1}{C{1.4cm}}{$\Delta$(CCSD(T):\newline UHF)} & \multicolumn{1}{C{1.4cm}}{$\Delta$(CCSD(T):\newline $\kappa$-OOMP2)} & \multicolumn{1}{C{1.4cm}}{$\Delta$(CCSD(T):\newline OOMP2)} & \multicolumn{1}{C{1.4cm}}{$\Delta$(CCSD(T):\newline BLYP)} & \multicolumn{1}{C{1.4cm}}{$\Delta$(CCSD(T):\newline B97M-rV)} & \multicolumn{1}{C{1.4cm}}{$\Delta$(CCSD(T):\newline B97)} & \multicolumn{1}{C{1.4cm}}{$\Delta$(CCSD(T):\newline$\omega$B97X-V)} & \multicolumn{1}{C{1.4cm}}{$\Delta$(CCSD(T):\newline$\omega$B97M-V)} \\
\hline
RMSD & 41.05   & 17.66  & 18.82  & 24.75   & 26.75   & 26.35   & 31.19   & 29.77   \\
MSD  & 4.35    & -2.19  & -1.80  & -4.99   & -4.36   & -4.05   & -4.71   & -5.45   \\
MIN  & -240.71 & -64.06 & -64.26 & -142.01 & -186.68 & -187.50 & -246.37 & -184.40 \\
MAX  & 177.65  & 90.68  & 88.43  & 87.87   & 86.24   & 88.25   & 89.26   & 89.52   \\
\hline
\end{tabular}
\end{table}

\subsection{Closed-shell subset}
\begin{landscape}
\tiny
\begin{ThreePartTable}
\begin{longtable}{L{1.2cm}L{1.2cm}L{1.2cm}R{1.4cm}R{1.4cm}R{1.5cm}R{1.4cm}R{1.4cm}R{1.4cm}R{1.4cm}R{1.4cm}R{1.5cm}}
\caption{Experimental vibrational frequencies (in cm$^{-1}$) and errors (in cm$^{-1}$) in the corrected vibrational frequencies for the 36 closed-shell species are presented in for the CCSD(T) methods utilizing different molecular orbitals. Root mean square deviations, mean signed deviations, most negative deviations, and most positive deviations (all in cm$^{-1}$) for the set of species and subsets are also presented.}\label{tab:closedshell} \\
\hline
Row 2--\newline Row 2 & Dimer & State & \multicolumn{1}{C{1.4cm}}{Expt.} & \multicolumn{1}{C{1.4cm}}{$\Delta$(CCSD(T):\newline UHF)} & \multicolumn{1}{C{1.5cm}}{$\Delta$(CCSD(T):\newline$\kappa$-OOMP2)} & \multicolumn{1}{C{1.4cm}}{$\Delta$(CCSD(T):\newline OOMP2)} & \multicolumn{1}{C{1.4cm}}{$\Delta$(CCSD(T):\newline BLYP)} & \multicolumn{1}{C{1.4cm}}{$\Delta$(CCSD(T):\newline B97M-rV)} & \multicolumn{1}{C{1.4cm}}{$\Delta$(CCSD(T):\newline B97)} & \multicolumn{1}{C{1.4cm}}{$\Delta$(CCSD(T):\newline $\omega$B97X-V)} & \multicolumn{1}{C{1.5cm}}{$\Delta$(CCSD(T):\newline $\omega$B97M-V)} \\
\cline{2-12} 
          & \ce{LiH}  & $X^{1}\Sigma^{+}$     & 1405.49805\tnote{a} & -7.12   & -8.81  & -2.59  & -8.41   & -9.83   & -2.50   & -8.68   & -6.13   \\
          & \ce{Li2}  & $X^{1}\Sigma_{g}^{+}$ & 351.4066\tnote{a}   & -11.40  & -16.69 & -16.52 & -15.37  & -13.54  & -1.76   & -16.24  & -11.40  \\
          & \ce{LiF}  & $X^{1}\Sigma^{+}$     & 910.57272\tnote{a}  & -5.10   & -5.60  & -6.23  & -6.96   & -6.37   & -6.23   & -6.09   & -6.07   \\
          & \ce{BeH+} & $X^{1}\Sigma^{+}$     & 2221.7\tnote{b}     & -11.81  & -13.69 & -12.35 & -16.41  & -13.93  & -16.22  & -11.43  & -10.61  \\
          & \ce{BeO}  & $X^{1}\Sigma^{+}$     & 1487.32\tnote{b}    & -20.79  & -17.36 & -21.83 & -20.41  & -19.22  & -20.00  & -18.16  & -18.40  \\
          & \ce{BH}   & $X^{1}\Sigma^{+}$     & 2366.7296\tnote{a}  & -2.51   & 0.49   & -1.09  & -1.00   & -1.63   & -1.02   & -0.97   & 2.51    \\
          & \ce{BF}   & $X^{1}\Sigma^{+}$     & 1402.15865\tnote{a} & -10.53 & -12.64  & -13.60 & -15.41  & -13.50  & -13.60  & -12.88  & -13.02  \\
          & \ce{C2}   & $X^{1}\Sigma_{g}^{+}$ & 1855.0663\tnote{a}  & 52.18   & -29.19 & -15.21 & -142.01 & -186.68 & -187.50 & -246.37 & -184.40 \\
          & \ce{CO}   & $X^{1}\Sigma^{+}$     & 2169.75589\tnote{a} & -16.15  & -18.53 & -19.25 & -20.34  & -18.52  & -18.76  & -18.10  & -18.13  \\
          & \ce{N2}   & $X^{1}\Sigma_{g}^{+}$ & 2358.57\tnote{a}    & -4.81   & -5.92  & -6.09  & -6.74   & -5.86   & -6.13   & -5.77   & -5.85   \\
          & \ce{NO+}  & $X^{1}\Sigma^{+}$     & 2376.72\tnote{a}    & -7.36   & -9.86  & -9.73  & -11.48  & -10.19  & -9.78   & -9.70   & -9.60 \\
          & \ce{OH-}  & $X^{1}\Sigma^{+}$     & 3735.2\tnote{c}     & -0.73   & -2.05  & -6.68  & -0.99   & 0.86    & -0.98   & 0.55    & -0.33  \\
          & \ce{HF}   & $X^{1}\Sigma^{+}$     & 4138.385\tnote{a}   & -7.04   & -9.66  & -8.96  & -8.06   & -7.40   & -8.24   & -7.48   & -7.40   \\
          & \ce{F2}   & $X^{1}\Sigma_{g}^{+}$ & 916.929\tnote{a}    & 60.05   & 3.10   & 3.57   & 3.20    & 3.06    & 3.00    & 2.93    & 2.73    \\
\cline{2-12}
          & RMSD      &                       &                     & 23.32   & 13.29  & 11.97  & 39.74   & 51.03   & 51.14   & 66.67   & 50.25   \\
          & MSD       &                       &                     & 0.49    & -10.46 & -9.75  & -19.31  & -21.62  & -20.69  & -25.60  & -20.44  \\
          & MIN       &                       &                     & -20.79  & -29.19 & -21.83 & -142.01 & -186.68 & -187.50 & -246.37 & -184.40 \\
          & MAX       &                       &                     & 60.05   & 3.10   & 3.57   & 3.20    & 3.06    & 3.00    & 2.93    & 2.73    \\
\hline 
Row 2--\newline Row 3 & Dimer & State & \multicolumn{1}{C{1.4cm}}{Expt.} & \multicolumn{1}{C{1.4cm}}{$\Delta$(CCSD(T):\newline UHF)} & \multicolumn{1}{C{1.5cm}}{$\Delta$(CCSD(T):\newline$\kappa$-OOMP2)} & \multicolumn{1}{C{1.4cm}}{$\Delta$(CCSD(T):\newline OOMP2)} & \multicolumn{1}{C{1.4cm}}{$\Delta$(CCSD(T):\newline BLYP)} & \multicolumn{1}{C{1.4cm}}{$\Delta$(CCSD(T):\newline B97M-rV)} & \multicolumn{1}{C{1.4cm}}{$\Delta$(CCSD(T):\newline B97)} & \multicolumn{1}{C{1.4cm}}{$\Delta$(CCSD(T):\newline $\omega$B97X-V)} & \multicolumn{1}{C{1.5cm}}{$\Delta$(CCSD(T):\newline $\omega$B97M-V)} \\
\cline{2-12}
          & \ce{NaH}  & $X^{1}\Sigma^{+}$     & 1171.968\tnote{a}   & -8.85   & -9.09  & -9.02  & 7.76    & -9.10   & 7.98    & 7.86    & 8.02    \\
          & \ce{NaLi} & $X^{1}\Sigma^{+}$     & 256.5412\tnote{a}   & -1.17   & -0.85  & -3.97  & -29.30  & -1.75   & -12.05  & -1.76   & -1.76   \\
          & \ce{NaF}  & $X^{1}\Sigma^{+}$     & 535.65805\tnote{a}  & -5.05   & -5.43  & -5.87  & -6.35   & -5.99   & -5.62   & -5.61   & -5.57   \\
          & \ce{MgH+} & $X^{1}\Sigma^{+}$     & 1699.1\tnote{b}     & -10.18  & -12.48 & -12.48 & -12.96  & -11.67  & -10.62  & -11.98  & -12.33  \\
          & \ce{AlH}  & $X^{1}\Sigma^{+}$     & 1682.37474\tnote{a} & -15.37  & -13.07 & -12.95 & -15.55  & -21.21  & -12.58  & -16.38  & -12.04  \\
          & \ce{AlF}  & $X^{1}\Sigma^{+}$     & 802.32447\tnote{a}  & -8.87   & -9.64  & -10.18 & -11.23  & -10.23  & -10.19  & -9.86   & -9.95   \\
          & \ce{SiH+} & $X^{1}\Sigma^{+}$     & 2157.17\tnote{a}    & -4.24   & -1.20  & -4.25  & -1.28   & -4.20   & -2.62   & -6.51   & -2.32   \\
          & \ce{SiO}  & $X^{1}\Sigma^{+}$     & 1241.54388\tnote{a} & -10.51  & -11.60 & -15.61 & -14.30  & -12.50  & -12.45  & -11.07  & -11.14  \\
          & \ce{PN}   & $X^{1}\Sigma^{+}$     & 1336.948\tnote{a}   & 33.01   & -5.90  & -7.60  & -6.56   & -5.54   & -5.84   & -5.45   & -5.57   \\
          & \ce{BeS}  & $X^{1}\Sigma^{+}$     & 997.94\tnote{a}     & -240.71 & -8.22  & -8.10  & -7.73   & -8.00   & -7.74   & -7.96   & -7.20   \\
          & \ce{CS}   & $X^{1}\Sigma^{+}$     & 1285.08\tnote{b}    & -5.98   & -7.54  & -9.54  & -9.31   & -8.00   & -7.98   & -6.94   & -7.13   \\
          & \ce{NS+}  & $X^{1}\Sigma^{+}$     & 1415\tnote{b}       & 52.10   & 3.37   & 3.04   & 2.40    & 3.51    & 3.15    & 3.47    & 3.24    \\
          & \ce{HCl}  & $X^{1}\Sigma^{+}$     & 2990.9248\tnote{a}  & 16.30   & 14.11  & 15.27  & 16.82   & 16.86   & 15.88   & 16.93   & 16.10   \\
          & \ce{LiCl} & $X^{1}\Sigma^{+}$     & 642.95453\tnote{a}  & -7.45   & -7.55  & -7.57  & -7.65   & -7.58   & -7.73   & -7.49   & -7.82   \\
          & \ce{BCl}  & $X^{1}\Sigma^{+}$     & 840.29472\tnote{a}  & -2.91   & -2.44  & -3.45  & -4.81   & -4.04   & -3.51   & -2.66   & -2.84   \\
          & \ce{CCl+} & $X^{1}\Sigma^{+}$     & 1175\tnote{b}       & 2.19    & -4.74  & -11.35 & -9.69   & -6.52   & -6.24   & -3.41   & -3.94   \\
          & \ce{ClF}  & $X^{1}\Sigma^{+}$     & 783.4534\tnote{a}   & -7.29   & -7.02  & -6.99  & -6.08   & -5.87   & -6.00   & -6.31   & -6.33   \\
\cline{2-12}
          & RMSD      &                       &                     & 60.79   & 8.31   & 9.47   & 11.86   & 9.68    & 8.89    & 8.83    & 8.23    \\
          & MSD       &                       &                     & -13.23  & -5.25  & -6.51  & -6.81   & -5.99   & -4.95   & -4.42   & -4.03   \\
          & MIN       &                       &                     & -240.71 & -13.07 & -15.61 & -29.30  & -21.21  & -12.58  & -16.38  & -12.33  \\
          & MAX       &                       &                     & 52.10   & 14.11  & 15.27  & 16.82   & 16.86   & 15.88   & 16.93   & 16.10   \\
\hline 
Row 3--\newline Row 3 & Dimer & State & \multicolumn{1}{C{1.4cm}}{Expt.} & \multicolumn{1}{C{1.4cm}}{$\Delta$(CCSD(T):\newline UHF)} & \multicolumn{1}{C{1.5cm}}{$\Delta$(CCSD(T):\newline$\kappa$-OOMP2)} & \multicolumn{1}{C{1.4cm}}{$\Delta$(CCSD(T):\newline OOMP2)} & \multicolumn{1}{C{1.4cm}}{$\Delta$(CCSD(T):\newline BLYP)} & \multicolumn{1}{C{1.4cm}}{$\Delta$(CCSD(T):\newline B97M-rV)} & \multicolumn{1}{C{1.4cm}}{$\Delta$(CCSD(T):\newline B97)} & \multicolumn{1}{C{1.4cm}}{$\Delta$(CCSD(T):\newline $\omega$B97X-V)} & \multicolumn{1}{C{1.5cm}}{$\Delta$(CCSD(T):\newline $\omega$B97M-V)} \\
\cline{2-12}
          & \ce{NaCl} & $X^{1}\Sigma^{+}$     & 364.6842\tnote{a}   & -5.25   & -5.27  & -5.29  & -5.31   & -5.41   & -5.31   & -5.27   & -5.29   \\
          & \ce{AlCl} & $X^{1}\Sigma^{+}$     & 481.77466\tnote{a}  & -7.43   & -7.38  & -7.50  & -7.88   & -7.70   & -7.60   & -7.44   & -7.47   \\
          & \ce{SiS}  & $X^{1}\Sigma^{+}$     & 749.64559\tnote{a}  & -8.84   & -8.35  & -9.41  & -9.62   & -9.05   & -8.96   & -8.42   & -8.43   \\
          & \ce{P2}   & $X^{1}\Sigma_{g}^{+}$ & 780.77\tnote{a}     & 32.77   & -4.41  & -4.67  & -4.68   & -4.38   & -4.48   & -4.47   & -4.52   \\
          & \ce{Cl2}  & $X^{1}\Sigma_{g}^{+}$ & 559.751\tnote{a}    & -11.83  & -11.64 & -11.69 & -11.64  & -11.63  & -11.59  & -11.66  & -11.63  \\
\cline{2-12}
          & RMSD      &                       &                     & 16.58   & 7.84   & 8.14   & 8.25    & 8.06    & 8.01    & 7.87    & 7.88    \\
          & MSD       &                       &                     & -0.11   & -7.41  & -7.71  & -7.83   & -7.64   & -7.59   & -7.45   & -7.47   \\
          & MIN       &                       &                     & -11.83  & -11.64 & -11.69 & -11.64  & -11.63  & -11.59  & -11.66  & -11.63  \\
          & MAX       &                       &                     & 32.77   & -4.41  & -4.67  & -4.68   & -4.38   & -4.48   & -4.47   & -4.52   \\
\hline 
Closed-\newline shell & & & & \multicolumn{1}{C{1.4cm}}{$\Delta$(CCSD(T):\newline UHF)} & \multicolumn{1}{C{1.5cm}}{$\Delta$(CCSD(T):\newline$\kappa$-OOMP2)} & \multicolumn{1}{C{1.4cm}}{$\Delta$(CCSD(T):\newline OOMP2)} & \multicolumn{1}{C{1.4cm}}{$\Delta$(CCSD(T):\newline BLYP)} & \multicolumn{1}{C{1.4cm}}{$\Delta$(CCSD(T):\newline B97M-rV)} & \multicolumn{1}{C{1.4cm}}{$\Delta$(CCSD(T):\newline B97)} & \multicolumn{1}{C{1.4cm}}{$\Delta$(CCSD(T):\newline $\omega$B97X-V)} & \multicolumn{1}{C{1.5cm}}{$\Delta$(CCSD(T):\newline $\omega$B97M-V)} \\
\cline{2-12}
          & RMSD      &                       &                     & 44.66   & 10.48  & 10.36  & 26.27   & 32.65   & 32.61   & 42.12   & 31.98   \\
          & MSD       &                       &                     & -6.07   & -7.58  & -7.94  & -11.82  & -12.30  & -11.44  & -13.08  & -10.89  \\
          & MIN       &                       &                     & -240.71 & -29.19 & -21.83 & -142.01 & -186.68 & -187.50 & -246.37 & -184.40 \\
          & MAX       &                       &                     & 60.05   & 14.11  & 15.27  & 16.82   & 16.86   & 15.88   & 16.93   & 16.10   \\
\hline
\end{longtable}
\begin{tablenotes}
\item[a] From Ref. \citenum{irikura2007experimental}.
\item[b] From Ref. \citenum{huber2018constants}.
\item[c] From Ref. \citenum{hotop1974high}.
\end{tablenotes}
\end{ThreePartTable}
\end{landscape}

Tab. \ref{tab:closedshell} presents the experimental vibrational frequencies, in cm$^{-1}$, and errors in the corrected vibrational frequencies, in cm$^{-1}$, for the 36 species in the closed-shell subset. We see that the use of OOMP2 or $\kappa$-OOMP2 orbitals is able to reduce the RMSD of UHF orbitals by a factor of 4 from 44.66 cm$^{-1}$ to 10.36 and 10.48 cm$^{-1}$, respectively. The use of DFT orbitals is also seen to reduce the RMSD, with BLYP outperforming the other functionals and $\omega$B97X-V improving the RMSD by only 2.5 cm$^{-1}$.  On average, all methods are seen to red-shift the closed-shell frequencies by 5-12 cm$^{-1}$. 

The performance of CCSD(T) with DFT orbitals is most negatively impacted for the \ce{C2} system, with errors in excess of -100 cm$^{-1}$ for each. This data point also represents a negative outlier for CCSD(T) with $\kappa$-OOMP2 orbitals and a positive outlier for CCSD(T) with UHF orbitals. Its significant MR character renders \ce{C2} outside of the scope of the single-reference methods evaluated in this work\cite{watts1992coupled,karton2011w4}.  
Judging from the mean-field $\langle S^{2}\rangle$ values (Tab. \ref{tab:ccsdt_s2}) for $\kappa$-OOMP2 and OOMP2 of 0.89140 and 0, respectively, we observe that OOMP2 is ``artificially" restoring symmetry in this case while the broken spin-symmetry of the $\kappa$-OOMP2 reference orbitals is diagnostic of a MR problem, consistent with the literature\cite{watts1992coupled,karton2011w4}. Spin-symmetry breaking in $\kappa$-OOMP2 as a signal of MR character has been demonstrated earlier in a study of fullerenes by Lee and Head-Gordon\cite{lee2019distinguishing}.   
Considering the other dimers in the closed-shell subset, CCSD(T) with UHF orbitals is seen to exhibit poor performance for \ce{F2}, \ce{PN}, \ce{BeS}, \ce{NS+}, and \ce{P2}, especially given that all other methods have absolute errors below 10 cm$^{-1}$ for these species. These errors can be attributed to spin contamination of the reference orbital set. For UHF, the mean field $\langle S^{2}\rangle$ values for \ce{F2}, \ce{PN}, \ce{BeS}, \ce{NS+}, and \ce{P2} are  0.31922, 0.70716, 1.01647, 0.69154, and 0.67604, respectively while the mean-field $\langle S^{2}\rangle$ values of $\kappa$-OOMP2, OOMP2, and the five DFT functionals for these species are all zero.  The large error for \ce{BeS} at the CCSD(T):UHF level can be traced to the character of the UHF wavefunction, which localizes nearly an entire electron's spin density on each atomic center. The $\langle S^{2}\rangle$ value of 1.01647 shows significant triplet contamination. In the absence of these symmetry-broken species (and MR \ce{C2}), the performance across the methods is largely equalized with RMSDs ranging from 9.5 to 12 cm$^{-1}$. The maximum error outlier for all non-HF references is attributed to HCl. \ce{NaLi} is also an outlying case for CCSD(T):BLYP (-29.30 cm$^{-1}$).

\subsection{Open-shell subset}
\begin{landscape}
\tiny
\begin{ThreePartTable}
\begin{longtable}{L{1.2cm}L{1.2cm}L{1.2cm}R{1.4cm}R{1.4cm}R{1.5cm}R{1.4cm}R{1.4cm}R{1.4cm}R{1.4cm}R{1.4cm}R{1.5cm}}
\caption{Experimental vibrational frequencies (in cm$^{-1}$) and errors (in cm$^{-1}$) in the corrected vibrational frequencies for the 59 open-shell species are presented in for the CCSD(T) methods utilizing different molecular orbitals. Root mean square deviations, mean signed deviations, most negative deviations, and most positive deviations (all in cm$^{-1}$) for the set of species and subsets are  presented.}\label{tab:openshell} \\
\hline
Row 2--\newline Row 2 & Dimer & State & \multicolumn{1}{C{1.4cm}}{Expt.} & \multicolumn{1}{C{1.4cm}}{$\Delta$(CCSD(T):\newline UHF)} & \multicolumn{1}{C{1.5cm}}{$\Delta$(CCSD(T):\newline $\kappa$-OOMP2)} & \multicolumn{1}{C{1.4cm}}{$\Delta$(CCSD(T):\newline OOMP2)} & \multicolumn{1}{C{1.4cm}}{$\Delta$(CCSD(T):\newline BLYP)} & \multicolumn{1}{C{1.4cm}}{$\Delta$(CCSD(T):\newline B97M-rV)} & \multicolumn{1}{C{1.4cm}}{$\Delta$(CCSD(T):\newline B97)} & \multicolumn{1}{C{1.4cm}}{$\Delta$(CCSD(T):\newline $\omega$B97X-V)} & \multicolumn{1}{C{1.5cm}}{$\Delta$(CCSD(T):\newline $\omega$B97M-V)} \\
\cline{2-12} 
          & \ce{LiO}   & $X^{2}\Pi_{i}$        & 814.62\tnote{a}     & -6.94   & -7.10   & -7.22   & -8.16   & -7.02   & -7.04   & -6.97   & -6.94   \\
          & \ce{BeH}   & $X^{2}\Sigma^{+}$     & 2061.235\tnote{a}   & -5.15   & -10.86  & -3.14   & -3.30   & -7.48   & -5.31   & -4.56   & -11.46  \\
          & \ce{BeF}   & $X^{2}\Sigma^{+}$     & 1247.36\tnote{b}    & 9.50    & 8.85    & 8.47    & 7.42    & 8.23    & 8.30    & 8.61    & 8.42    \\
          & \ce{B2}    & $X^{3}\Sigma_{g}^{-}$ & 1051.3\tnote{b}     & -16.10  & -29.18  & -1.90   & -90.13  & -72.72  & -61.76  & -45.34  & -75.26  \\
          & \ce{BN}    & $X^{3}\Pi$            & 1514.6\tnote{b}     & 6.82    & -8.55   & -7.84   & -7.27   & -7.70   & -7.12   & -7.95   & -8.55   \\
          & \ce{BO}    & $X^{2}\Sigma^{+}$     & 1885.286\tnote{a}   & -11.61  & -21.02  & -23.17  & -24.32  & -21.41  & -22.38  & -21.98  & -21.92  \\
          & \ce{CH}    & $X^{2}\Pi_{r}$        & 2860.7508\tnote{a}  & -10.43  & 1.87    & -1.66   & -1.56   & -0.88   & -0.74   & -0.33   & -2.12   \\
          & \ce{C2-}   & $X^{2}\Sigma_{g}^{+}$ & 1781.189\tnote{a}   & -116.46 & -6.89   & -8.56   & -7.90   & -6.57   & -4.75   & -46.93  & -133.20 \\
          & \ce{CN}    & $X^{2}\Sigma^{+}$     & 2068.648\tnote{a}   & 54.67   & -7.85   & -10.32  & -8.92   & -7.36   & -7.65   & -7.05   & -7.37   \\
          & \ce{CO+}   & $X^{2}\Sigma^{+}$     & 2214.127\tnote{a}   & 67.00   & -16.90  & -20.40  & -19.77  & -13.55  & -16.29  & -16.28  & -16.00  \\
          & \ce{CF}    & $X^{2}\Pi_{r}$        & 1307.93\tnote{a}    & -3.10   & -6.70   & -8.59   & -10.31  & -7.35   & -7.36   & -6.36   & -6.52   \\
          & \ce{NH}    & $X^{3}\Sigma^{-}$     & 3282.72\tnote{a}    & 0.19    & -1.75   & -0.73   & -1.66   & -1.85   & -0.93   & -0.68   & -0.93   \\
          & \ce{N2+}   & $X^{2}\Sigma_{g}^{+}$ & 2207.0115\tnote{a}  & 70.77   & -6.08   & -7.18   & -6.92   & -4.96   & -5.51   & -4.79   & -5.09   \\
          & \ce{NO}    & $X^{2}\Pi_{r}$        & 1904.1346\tnote{a}  & 177.65  & -6.77   & -6.31   & -7.80   & -6.83   & -7.09   & -6.89   & -6.90   \\
          & \ce{NF}    & $X^{3}\Sigma^{-}$     & 1141.37\tnote{a}    & -7.98   & -10.86  & -13.45  & -14.29  & -9.97   & -10.40  & -9.82   & -9.56   \\
          & \ce{OH}    & $X^{2}\Pi_{i}$        & 3737.761\tnote{a}   & -10.03  & -9.36   & -10.91  & -11.21  & -9.13   & -9.16   & -8.56   & -10.23  \\
          & \ce{OH+} & $X^{3}\Sigma^{-}$ & 3113.37\tnote{b} & 6.88 & 4.44 & 4.52 & 6.84 & 6.76 & 4.34 & 4.44 & 4.32 \\
          & \ce{O2}    & $X^{3}\Sigma_{g}^{-}$ & 1580.161\tnote{a}   & -4.57   & -3.09   & -3.11   & -3.25   & -3.05   & -3.23   & -3.22   & -3.24   \\
          & \ce{O2+}   & $X^{2}\Pi_{g}$        & 1905.892\tnote{a}   & 28.13   & 4.68    & 4.77    & 3.87    & 4.66    & 4.40    & 4.54    & 4.45    \\
          & \ce{O2-}   & $X^{2}\Pi_{g,i}$      & 1090\tnote{b}       & 24.81   & 27.20   & 27.23   & 27.20   & 27.12   & 26.89   & 26.82   & 26.85   \\
          & \ce{OF}    & $X^{2}\Pi$            & 1053.0138\tnote{a}  & 29.52   & 2.56    & 0.62    & 1.23    & 5.31    & 4.65    & 5.21    & 5.54    \\
          & \ce{HF+}   & $X^{2}\Pi_{i}$        & 3090.5\tnote{b}     & 26.49   & 27.97   & 29.45   & 27.38   & 28.07   & 26.44   & 26.44   & 26.78   \\
           & \ce{F2+} & $X^{2}\Pi_{g,i}$ & 1104\tnote{d} & -34.13 & 20.51 & 23.81 & 19.99 & 19.30 & 18.63 & 17.94 & 17.71 \\
          & \ce{F2-}   & $X^{2}\Sigma_{u}^{+}$ & 510\tnote{b*}        & -40.09  & -64.06  & -64.26  & -63.25  & -64.22  & -63.46  & -63.22  & -63.09  \\
\cline{2-12}
 & RMSD &  &  & 51.77   & 18.77  & 18.44  & 25.78  & 23.00  & 21.46  & 21.76  & 35.74   \\
 & MSD  &  &  & 9.83    & -4.96  & -4.16  & -8.17  & -6.36  & -6.11  & -6.96  & -12.26  \\
 & MIN  &  &  & -116.46 & -64.06 & -64.26 & -90.13 & -72.72 & -63.46 & -63.22 & -133.20 \\
 & MAX  &  &  & 177.65  & 27.97  & 29.45  & 27.38  & 28.07  & 26.89  & 26.82  & 26.85  \\
\hline 
Row 2--\newline Row 3 & Dimer & State & \multicolumn{1}{C{1.4cm}}{Expt.} & \multicolumn{1}{C{1.4cm}}{$\Delta$(CCSD(T):\newline UHF)} & \multicolumn{1}{C{1.4cm}}{$\Delta$(CCSD(T):\newline$\kappa$-OOMP2)} & \multicolumn{1}{C{1.4cm}}{$\Delta$(CCSD(T):\newline OOMP2)} & \multicolumn{1}{C{1.4cm}}{$\Delta$(CCSD(T):\newline BLYP)} & \multicolumn{1}{C{1.4cm}}{$\Delta$(CCSD(T):\newline B97M-rV)} & \multicolumn{1}{C{1.4cm}}{$\Delta$(CCSD(T):\newline B97)} & \multicolumn{1}{C{1.4cm}}{$\Delta$(CCSD(T):\newline $\omega$B97X-V)} & \multicolumn{1}{C{1.4cm}}{$\Delta$(CCSD(T):\newline $\omega$B97M-V)} \\
\cline{2-12} 
          & \ce{NaO}   & $X^{2}\Pi$            & 526\tnote{b*}        & -38.60  & -38.52  & -38.69  & -38.45  & -38.46  & -38.58  & -38.44  & -38.59  \\
          & \ce{MgH}   & $X^{2}\Sigma^{+}$     & 1492.7763\tnote{a}  & -4.27   & -2.49   & -4.24   & -5.69   & -2.78   & -0.51   & -3.21   & -4.82   \\
          & \ce{MgF}   & $X^{2}\Sigma^{+}$     & 711.69\tnote{b}     & 0.08    & -0.30   & -0.66   & -1.34   & -0.93   & -0.78   & -0.50   & -0.48   \\
          & \ce{AlH+}  & $X^{2}\Sigma^{+}$     & 1620\tnote{b}       & 24.02   & 29.86   & 21.80   & 34.31   & 28.10   & 28.30   & 27.77   & 19.42   \\
          & \ce{SiH}   & $X^{2}\Pi_{r}$        & 2042.5229\tnote{a}  & -12.95  & -9.14   & -9.12   & -9.46   & -7.06   & -7.05   & -12.62  & -12.63  \\
          & \ce{SiF}   & $X^{2}\Pi_{r}$        & 837.32507\tnote{a}  & 10.84   & 9.61    & 8.85    & 7.61    & 8.75    & 8.85    & 9.39    & 9.29    \\
          & \ce{PH}    & $X^{3}\Sigma^{-}$     & 2363.774\tnote{a}   & 1.72    & 0.54    & -1.71   & 1.43    & -0.52   & 1.70    & 1.47    & 1.82    \\
          & \ce{PH+}   & $X^{2}\Pi_{r}$        & 2299.6\tnote{b}     & 90.81   & 90.68   & 88.43   & 87.87   & 86.24   & 88.25   & 89.26   & 89.52   \\
          & \ce{PH-}   & $X^{2}\Pi_{i}$        & 2230\tnote{b}       & 30.09   & 31.71   & 33.09   & 27.12   & 33.24   & 24.68   & 31.71   & 31.07   \\
          & \ce{CP}    & $X^{2}\Sigma^{+}$     & 1239.79924\tnote{a} & 35.49   & 4.78    & -7.69   & -4.50   & 1.77    & 0.64    & 6.27    & 1.97    \\
          & \ce{PO}    & $X^{2}\Pi_{r}$        & 1233.34\tnote{a}    & -3.47   & -6.44   & -8.11   & -6.74   & -5.69   & -5.73   & -5.24   & -5.33   \\
          & \ce{PO-}   & $X^{3}\Sigma^{-}$     & 1000\tnote{b}       & 34.68   & 32.77   & 30.68   & 33.46   & 34.26   & 34.26   & 34.36   & 34.19   \\
          & \ce{PF}    & $X^{3}\Sigma^{-}$     & 846.75\tnote{a}     & -7.52   & -8.36   & -9.11   & -10.09  & -8.75   & -8.78   & -8.38   & -8.48   \\
          & \ce{PF+}   & $X^{2}\Pi_{r}$        & 1053.25\tnote{b}    & -5.85   & -7.91   & -9.70   & -11.04  & -8.82   & -8.66   & -7.70   & -7.83   \\
          & \ce{HS}    & $X^{2}\Pi_{i}$        & 2696.2475\tnote{a}  & 7.92    & 8.60    & 8.60    & 8.59    & 8.56    & 8.58    & 8.57    & 6.39    \\
          & \ce{BS}    & $X^{2}\Sigma^{+}$     & 1179.91\tnote{a}    & 1.85    & -2.78   & -4.02   & -4.42   & -3.32   & -3.49   & -3.09   & -3.21   \\
          & \ce{CS+}   & $X^{2}\Sigma^{+}$     & 1384\tnote{b}       & 30.28   & -3.84   & -13.40  & -10.72  & -2.47   & -5.36   & -2.25   & -4.66   \\
          & \ce{NS}    & $X^{2}\Pi_{r}$        & 1218.7\tnote{b}     & 24.33   & -3.17   & -3.59   & -4.01   & -2.91   & -2.67   & -1.69   & -2.36   \\
          & \ce{SO}    & $X^{3}\Sigma^{-}$     & 1150.7913\tnote{a}  & 2.02    & -1.02   & -2.19   & 0.40    & 1.28    & 0.91    & 0.86    & 0.82    \\
          & \ce{SO+}   & $X^{2}\Pi_{r}$        & 1306.778\tnote{e}  & 21.21   & 0.55    & -0.43   & 1.81    & 2.81    & 2.61    & 2.85    & 2.76    \\
          & \ce{HCl+}  & $X^{2}\Pi_{i}$        & 2673.69\tnote{a}    & 26.52   & 23.76   & 23.84   & 23.45   & 24.90   & 23.34   & 24.97   & 23.78   \\
          & \ce{LiCl-} & $X^{2}\Sigma^{+}$     & 480\tnote{b}        & 26.79   & 27.15   & 26.71   & 27.08   & 27.11   & 26.71   & 27.15   & 27.04   \\
          & \ce{BeCl}  & $X^{2}\Sigma^{+}$     & 846.7\tnote{b}      & -2.93   & -2.97   & -3.03   & -3.26   & -3.30   & -3.10   & -2.98   & -3.02   \\
          & \ce{CCl}   & $X^{2}\Pi$            & 876.89749\tnote{a}  & -6.99   & -7.55   & -10.06  & -10.92  & -9.15   & -9.01   & -7.69   & -7.96   \\
          & \ce{NCl}   & $X^{3}\Sigma^{-}$     & 827.95767\tnote{a}  & -11.14  & -10.50  & -12.85  & -13.73  & -11.79  & -11.82  & -10.95  & -10.81  \\
          & \ce{OCl}   & $X^{2}\Pi_{i}$        & 853.64268\tnote{a}  & -5.50   & -4.64   & -4.10   & -4.86   & -5.24   & -5.15   & -4.87   & -4.87   \\
          & \ce{ClF+}  & $X^{2}\Pi$            & 870\tnote{b}        & 50.44   & 45.33   & 45.39   & 47.50   & 47.73   & 47.24   & 46.76   & 46.72   \\
\cline{2-12} 
          & RMSD       &                       &                     & 27.42   & 24.88   & 24.52   & 24.95   & 24.58   & 24.38   & 24.88   & 24.52   \\
          & MSD        &                       &                     & 11.85   & 7.25    & 5.36    & 5.98    & 7.17    & 6.87    & 7.47    & 6.66    \\
          & MIN        &                       &                     & -38.60  & -38.52  & -38.69  & -38.45  & -38.46  & -38.58  & -38.44  & -38.59  \\
          & MAX        &                       &                     & 90.81   & 90.68   & 88.43   & 87.87   & 86.24   & 88.25   & 89.26   & 89.52   \\
\hline
Row 3-Row 3 & Dimer & State & \multicolumn{1}{C{1.4cm}}{Expt.} & \multicolumn{1}{C{1.4cm}}{$\Delta$(CCSD(T):\newline UHF)} & \multicolumn{1}{C{1.5cm}}{$\Delta$(CCSD(T):\newline$\kappa$-OOMP2)} & \multicolumn{1}{C{1.4cm}}{$\Delta$(CCSD(T):\newline OOMP2)} & \multicolumn{1}{C{1.4cm}}{$\Delta$(CCSD(T):\newline BLYP)} & \multicolumn{1}{C{1.4cm}}{$\Delta$(CCSD(T):\newline B97M-rV)} & \multicolumn{1}{C{1.4cm}}{$\Delta$(CCSD(T):\newline B97)} & \multicolumn{1}{C{1.4cm}}{$\Delta$(CCSD(T):\newline $\omega$B97X-V)} & \multicolumn{1}{C{1.5cm}}{$\Delta$(CCSD(T):\newline $\omega$B97M-V)} \\
\cline{2-12}
          & \ce{MgCl}  & $X^{2}\Sigma^{+}$     & 462.12\tnote{b}     & 2.24    & 2.27    & 2.27    & 2.14    & 2.19    & 2.27    & 2.24    & 2.31    \\
          & \ce{AlS}   & $X^{2}\Sigma^{+}$     & 617.1169\tnote{a}   & -8.83   & -14.00  & -16.97  & -15.43  & -15.34  & -14.91  & -13.45  & -13.33  \\
          & \ce{Si2}   & $X^{3}\Sigma_{g}^{-}$ & 510.98\tnote{a}     & -0.03   & 0.26    & 33.72   & 0.50    & 0.46    & 0.52    & 0.32    & 0.34    \\
          & \ce{SiCl}  & $X^{2}\Pi_{r}$        & 535.59\tnote{a}     & -6.92   & -6.88   & -7.09   & -7.63   & -7.34   & -7.19   & -6.83   & -6.91   \\
          & \ce{P2+}   & $X^{2}\Pi_{u}$        & 672.2\tnote{a}      & 33.88   & 8.27    & 60.95   & 8.56    & 8.68    & 8.56    & 8.25    & 8.27    \\
          & \ce{PS}    & $X^{2}\Pi_{r}$        & 739.1\tnote{b}      & 19.14   & -5.46   & -5.93   & -5.94   & -5.73   & -5.64   & -5.81   & -5.71   \\
          & \ce{S2}    & $X^{3}\Sigma_{g}^{-}$ & 725.7102\tnote{a}   & -7.69   & -7.18   & -7.44   & -7.06   & -7.05   & -7.02   & -7.17   & -7.13   \\
          & \ce{S2+}   & $X^{2}\Pi_{g,r}$      & 790\tnote{b}        & 44.58   & 11.06   & 10.91   & 11.14   & 11.32   & 11.25   & 11.06   & 11.07   \\
\cline{2-12}
          & RMSD       &                       &                     & 21.48   & 8.07    & 25.99   & 8.54    & 8.52    & 8.37    & 7.97    & 7.95    \\
          & MSD        &                       &                     & 9.55    & -1.46   & 8.80    & -1.71   & -1.60   & -1.52   & -1.42   & -1.39   \\
          & MIN        &                       &                     & -8.83   & -14.00  & -16.97  & -15.43  & -15.34  & -14.91  & -13.45  & -13.33  \\
          & MAX        &                       &                     & 44.58   & 11.06   & 60.95   & 11.14   & 11.32   & 11.25   & 11.06   & 11.07   \\
\hline 
Open-\newline shell & & & & \multicolumn{1}{C{1.4cm}}{$\Delta$(CCSD(T):\newline UHF)} & \multicolumn{1}{C{1.5cm}}{$\Delta$(CCSD(T):\newline$\kappa$-OOMP2)} & \multicolumn{1}{C{1.4cm}}{$\Delta$(CCSD(T):\newline OOMP2)} & \multicolumn{1}{C{1.4cm}}{$\Delta$(CCSD(T):\newline BLYP)} & \multicolumn{1}{C{1.4cm}}{$\Delta$(CCSD(T):\newline B97M-rV)} & \multicolumn{1}{C{1.4cm}}{$\Delta$(CCSD(T):\newline B97)} & \multicolumn{1}{C{1.4cm}}{$\Delta$(CCSD(T):\newline $\omega$B97X-V)} & \multicolumn{1}{C{1.5cm}}{$\Delta$(CCSD(T):\newline $\omega$B97M-V)} \\
\cline{2-12}
& RMSD &  &  & 38.69   & 20.87  & 22.47  & 23.77  & 22.39  & 21.65  & 22.01  & 28.34   \\
& MSD  &  &  & 10.71   & 1.10   & 1.95   & -0.82  & 0.48   & 0.45   & 0.40   & -2.13   \\
& MIN  &  &  & -116.46 & -64.06 & -64.26 & -90.13 & -72.72 & -63.46 & -63.22 & -133.20 \\
& MAX  &  &  & 177.65  & 90.68  & 88.43  & 87.87  & 86.24  & 88.25  & 89.26  & 89.52    \\
\hline
\end{longtable}
\begin{tablenotes}
\item[a] From Ref. \citenum{irikura2007experimental}.
\item[b] From Ref. \citenum{huber2018constants}.
\item[d] From Ref. \citenum{cormack1996high}.
\item[e] From Ref. \citenum{milkman1988high}.
\item[*] Theoretical results.
\end{tablenotes}
\end{ThreePartTable}
\end{landscape}

Tab. \ref{tab:openshell} presents the experimental vibrational frequencies, in cm$^{-1}$, and errors in the corrected vibrational frequencies, in cm$^{-1}$, for the 59 species in the open-shell subset. Turning to Fig. \ref{fig:boxplot_all}, the open-shell non-HF methods exhibit an increase in the number of outliers as compared to the closed-shell cases. 
Overall, CCSD(T) with $\kappa$-OOMP2 orbitals provides the best RMSD of all references (20.87 cm$^{-1}$), improving on the performance of CCSD(T) with UHF orbitals by nearly a factor of two (38.69 cm$^{-1}$). The use of OOMP2, BLYP, B97M-rV, B97, and $\omega$B97X-V orbitals yields comparable performance to $\kappa$-OOMP2 orbitals (22-24 cm$^{-1}$), while the overall performance of CCSD(T) with $\omega$B97M-V orbitals falls between the other non-UHF references and the UHF reference (28.34 cm$^{-1}$). In terms of MSDs, CCSD(T) with UHF orbitals is seen to blue shift the open-shell frequencies by 11 cm$^{-1}$ while the non-UHF methods yield little-to-no systemic shift in frequencies ($\pm$ 2 cm$^{-1}$). The performance of CCSD(T) with $\kappa$-OOMP2 or OOMP2 orbitals on the open-shell systems is a factor of two worse than for the closed-shell systems. 

Looking at individual cases, errors in corrected vibrational frequencies for the triplet ground state of \ce{B2} are seen to range from -1.90 cm$^{-1}$ using OOMP2 orbitals to -90.13 cm$^{-1}$ using BLYP orbitals. \ce{B2} is another system which is know to exhibit MR character and therefore the varied and often poor performance of these single reference methods is to be expected\cite{karton2011w4}. Evidence of MR character is seen in the mean-field $\langle S^2 \rangle$ values, where the $\kappa$-OOMP2 and UHF orbitals both significantly break spin-symmetry ($\langle S^2 \rangle$ of	2.81206 and 2.90778, respectively; Tab. \ref{tab:ccsdt_s2}). The DFT orbital references are also seen to significantly break spin-symmetry while the OOMP2 reference artificially restores spin-symmetry for this system.

In agreement with the work of Beran et al.\cite{beran2003approaching} and Tentscher and Arey\cite{tentscher2012geometries} the predictions of the frequencies of \ce{CN}, \ce{NO}, \ce{OF}, and their isoelectronic and isovalent counterparts with CCSD(T) using a UHF reference yield sizeable errors. \ce{CN} is isoelectronic to \ce{C2-}, \ce{CO+}, and \ce{N2+} and is isovalent to \ce{CP}, \ce{CS+}, and \ce{P2+}. CCSD(T) with UHF orbitals yields errors in the corrected frequencies of these species of 54.67 cm$^{-1}$, -116.46 cm$^{-1}$, 67.00 cm$^{-1}$, 70.77 cm$^{-1}$, 35.49 cm$^{-1}$,  30.28 cm$^{-1}$, and 33.88 cm$^{-1}$, respectively, while CCSD(T) with $\kappa$-OOMP2 orbitals improves these errors to -7.85 cm$^{-1}$, -6.89 cm$^{-1}$, -16.90 cm$^{-1}$, -6.08 cm$^{-1}$, 4.78 cm$^{-1}$, -3.84 cm$^{-1}$, and 8.27 cm$^{-1}$, respectively. In all of these cases but \ce{C2-} these errors with UHF orbital references are accompanied by spin-symmetry breaking at the level of the reference with respective $\langle S^{2}\rangle$ values of 1.15755, 0.75627, 0.96970, 1.23889, 1.61297, 1.47259, 1.16866 for UHF and 0.76257, 0.75579, 0.76730, 0.75334, 0.82118, 0.80770, 0.75476 for the $\kappa$-OOMP2 reference (Tab. \ref{tab:ccsdt_s2}). The spin-symmetry restoration from the $\kappa$-OOMP2 (or OOMP2/DFT) orbitals is seen to dramatically improve the predicted frequencies for these systems with significant symmetry-breaking occurring at the UHF level. For \ce{C2-}, the error in predicted frequency for CCSD(T) with UHF orbitals has the opposite sign of the other frequency errors in this isoelectronic/isovalent family of dimers and the UHF $\langle S^{2}\rangle$ suggests little spin-contamination. Instead, UHF, $\omega$B97X-V, and $\omega$B97M-V are seen to favor broken-spatial-symmetry solutions for \ce{C2-}, contributing to large errors in the predicted frequencies while $\kappa$-OOMP2, OOMP2, and the other density functionals preserve the spatial symmetry and yield much more reliable frequencies.
For \ce{P2+}, CCSD(T):OOMP2 yields an error of 60.95 cm$^{-1}$, almost twice that of CCSD(T):UHF. The orbital optimization of the ground state at the OOMP2 level is shown to give preference to a higher symmetry orbital occupation where the $\pi_{3p_{x}}$ and $\pi_{3p_{y}}$ MOs are doubly-occupied and the $\sigma_{3p_{z}}$ orbital is singly-occupied. The ground state for all other methods doubly occupies the $\sigma_{3p_{z}}$ and singly-occupies one of the two $\pi_{3p}$ orbitals, breaking the $D_{\infty h}$ symmetry of the molecule. The latter occupation, however, is the filling predicted by MO theory and yields reasonable frequencies in comparison to the experimental benchmark. This represents another example of essential symmetry breaking that is quelled by OOMP2. 

\ce{NO} is isoelectronic to \ce{O2+} and isovalent to \ce{PO}, \ce{NS}, \ce{SO+}, \ce{PS}, and \ce{S2+}. CCSD(T) with UHF orbitals yields errors in the corrected frequencies of these species of 177.65 cm$^{-1}$, 28.13 cm$^{-1}$, -3.47 cm$^{-1}$, 24.33 cm$^{-1}$, 21.21 cm$^{-1}$, 19.14 cm$^{-1}$, and 44.58 cm$^{-1}$, respectively, while CCSD(T) with $\kappa$-OOMP2 orbitals yields error of -6.77 cm$^{-1}$, 4.68 cm$^{-1}$, -6.44 cm$^{-1}$, -3.17 cm$^{-1}$, 0.55 cm$^{-1}$, -5.46 cm$^{-1}$, and 11.06 cm$^{-1}$, respectively. \ce{NO} represents the most positive outlier for CCSD(T) with UHF orbitals while the error for \ce{PO} is the smallest of all the orbital references considered. Comparing the mean-field $\langle S^{2}\rangle$ values between UHF and $\kappa$-OOMP2 for these species, \ce{O2+} (1.12597 vs. 0.75272), \ce{NS} (1.18951	vs. 0.75758), \ce{SO+} (1.18709 vs. 0.75583), \ce{PS} (1.05380 vs. 0.76119), and \ce{S2+} (1.23007 vs 0.75862) show significant spin-contamination at the UHF level while cases exhibiting the largest and smallest errors, \ce{NO} (0.79621 vs. 0.75407) and \ce{PO} (0.77298 vs. 0.75758), respectively, show little evidence of significant spin-contamination (Tab. \ref{tab:ccsdt_s2}). These findings are consistent with those of Szalay et al.\cite{szalay2004triplet}, which find that in these errors arise from instabilities in the doublet wavefuction. This discrepancy between \ce{NO} and \ce{PO} arises due to the instability in the \ce{PO} UHF wavefunction occurring at a larger internuclear separation than the equilibrium bond length around which data was collected. 

\ce{OF} is isoelectronic to \ce{F2+} and isovalent to \ce{OCl} and \ce{ClF+}. CCSD(T) with UHF orbitals yields errors in the corrected frequencies of these species of 29.52 cm$^{-1}$, -21.63 cm$^{-1}$, -5.50 cm$^{-1}$, and 50.44 cm$^{-1}$, respectively, while CCSD(T) with $\kappa$-OOMP2 orbitals yields error of 2.56 cm$^{-1}$, 33.01 cm$^{-1}$, -4.64 cm$^{-1}$, and 45.33 cm$^{-1}$, respectively. The errors for UHF and non-UHF orbitals for \ce{F2+} are seen to differ by over 50 cm$^{-1}$ while for \ce{OCl} and \ce{ClF} the errors in the frequencies are in good agreement. Similarly to the cases of \ce{NO} and \ce{PO}, the UHF wavefunctions for \ce{OF}, \ce{OCl}, and \ce{ClF+} do not demonstrate signs of significant spin contamination ($\langle S^{2}\rangle$ values of 0.77257, 0.77010, 0.76463, respectively; Tab. \ref{tab:ccsdt_s2}). 

\subsection{Potential issues with experimental data}
\begin{ThreePartTable}
\tiny
\begin{longtable}{L{2.2cm}L{2.2cm}R{2.2cm}R{2.2cm}R{2.2cm}R{2.2cm}}
\caption{Experimental vibrational frequencies (in cm$^{-1}$), mean errors for the non-HF CCSD(T) methods (in cm$^{-1}$), ranges of the errors for the non-HF CCSD(T) methods (in cm$^{-1}$), and alternative experimental reported frequencies (in cm$^{-1}$) are presented for species where the experimental results are in question.}\label{tab:questionable} \\
\hline
Dimer & State & \multicolumn{1}{C{2.2cm}}{Current Expt.\newline Frequency} & \multicolumn{1}{C{2.2cm}}{Mean Non-HF\newline Error} & \multicolumn{1}{C{2.2cm}}{Non-HF Error\newline Range} & \multicolumn{1}{C{2.2cm}}{Alternative Expt. \newline Frequency} \\
\hline
\ce{O2-}   & $X^{2}\Pi_{g,i}$      & 1090\cite{huber2018constants}     & 27.04  & 0.41  &  1090\cite{boness1970structure,linder1971experimental} \\ 
& & & & & 1108 $\pm$ 20\cite{ervin2003only} \\ 
& & & & & 1140\cite{gray1972vibrational} \\ 
& & & & & 1145\cite{creighton1964vibrational} \\
\ce{HF+}   & $X^{2}\Pi_{i}$        & 3090.5\cite{huber2018constants,gewurtz1975electronic,hovde1989velocity}   & 27.51  & 3.01  & 3061.8\cite{yencha1999threshold} \\
& & & & & 3118\cite{beran2003approaching}\tnote{*} \\
& & & & & 3119\cite{tentscher2012geometries}\tnote{*} \\
\ce{F2+}   & $X^{2}\Pi_{g,i}$      & 1104\cite{cormack1996high} & 19.70  & 6.09  & 1091.5\cite{yang2005combined} \\
\ce{F2-}   & $X^{2}\Sigma_{u}^{+}$ & 510\cite{huber2018constants,gilbert1971single}\tnote{*}      & -63.65 & 1.17  &  \\
\ce{NaO}   & $X^{2}\Pi$            & 526\cite{huber2018constants}\tnote{*}      & -38.53 & 0.25  & 504\cite{o1972thermochemical}\tnote{*} \\
& & & & & 547\tnote{*}\cite{so1975electronic} \\
\ce{AlH+}  & $X^{2}\Sigma^{+}$     & 1620\cite{huber2018constants,holst1934bandenspektrum}     & 27.08  & 14.89 &  \\
\ce{PH+}   & $X^{2}\Pi_{r}$        & 2299.6\cite{huber2018constants,narasimham1957emission}   & 88.61  & 4.44  & 2382.75\cite{reddy1995dissociation}\tnote{*} \\
\ce{PH-}   & $X^{2}\Pi_{i}$        & 2230\cite{huber2018constants}     & 30.37  & 8.56  & 2230 $\pm$ 100\cite{zittel1976laser} \\
\ce{PO-}   & $X^{3}\Sigma^{-}$     & 1000\cite{huber2018constants}     & 33.43  & 3.68  & 1000 $\pm$ 70\cite{zittel1976laser} \\
\ce{HCl+}  & $X^{2}\Pi_{i}$        & 2673.69\cite{irikura2007experimental}  & 24.00  & 1.63  & 2702.6\cite{patanen2014high}\tnote{*} \\
\ce{LiCl-} & $X^{2}\Sigma^{+}$     & 480\cite{huber2018constants}      & 26.99  & 0.44  & 480 $\pm$ 80\cite{carlsten1976binding} \\
\ce{ClF+}  & $X^{2}\Pi$            & 870\cite{huber2018constants}      & 46.67  & 2.40  &  870 $\pm$ 30\cite{dekock1972photoelectron} \\
& & & & & 912 $\pm$ 30\cite{anderson1971photoelectron} \\
\hline
\end{longtable}
\begin{tablenotes}
\item[*] Theoretical results.
\end{tablenotes}
\end{ThreePartTable}
In the cases of \ce{ClF+} and \ce{PH+} the predicted frequencies for CCSD(T) with a UHF orbital reference and CCSD(T) with non-UHF orbital references differ significantly (more than 25 cm$^{-1}$) from the reported benchmark value while the CCSD(T) with non-UHF orbitals all yield predicted frequencies in agreement with each other. This observation is true of other molecules and ions in the open-shell set as well, as summarized in Tab. \ref{tab:questionable}. In all of these cases except \ce{F2+} and \ce{F2-} the errors for CCSD(T) with UHF orbitals are consistent with the errors for CCSD(T) with non-HF orbitals. The agreement of all methods suggests that potentially inaccurate or imprecise experimental reference values should be revisited. For \ce{NaO}\cite{o1972thermochemical,so1975electronic} and \ce{F2-}\cite{gilbert1971single}, the reference data given by Huber and Herzberg\cite{huber2018constants} are sourced from calculations performed at the HF level of theory; any of the CCSD(T) methods surveyed should be seen as a more accurate result for these systems. For another subset of these systems the error bars on the experimental values contain the frequencies calculated from this work (\ce{PH-}: 2230 $\pm$ 100 cm$^{-1}$\cite{zittel1976laser}, \ce{PO-}: 1000 $\pm$ 70 cm$^{-1}$\cite{zittel1976laser}, \ce{LiCl-}: 480 $\pm$ 80 cm$^{-1}$\cite{carlsten1976binding}). For  \ce{ClF+}, Huber and Herzberg\cite{huber2018constants} reference the  DeKock et al.\cite{dekock1972photoelectron} who report a vibrational frequency of 870 $\pm$ 30 cm$^{-1}$ while Anderson et al. \cite{anderson1971photoelectron} report a frequency of 912 $\pm$ 30 cm$^{-1}$. Our calculations, which predict a \ce{ClF+} frequency of 917 cm$^{-1}$, more closely agree with the work of Anderson et al. Similarly, for \ce{O2-} Huber and Herzberg\cite{huber2018constants} cite a value of 1090 cm$^{-1}$ based on the works of Boness and Schulz\cite{boness1970structure} and  Linder and Schmidt\cite{linder1971experimental} while also noting that Gray et al.\cite{gray1972vibrational} and Creighton and Lippincott\cite{creighton1964vibrational} give values of 1140 cm$^{-1}$ and 1145 cm$^{-1}$, respectively. In a more recent study, Ervin et al.\cite{ervin2003only} give a value of 1108 $\pm$ 20 cm$^{-1}$, which is in good agreement with our theoretical predictions of 1117 cm$^{-1}$. The most positive error point for many of the CCSD(T) method with non-UHF references, \ce{PH+}, does not have much experimental data on its spectroscopic constants in the literature, with the value of 2299.6 cm$^{-1}$ tracing back to a study by Narasimham\cite{narasimham1957emission}. A recent modeling study by Reddy et al.\cite{reddy1995dissociation} predicts a ground-state frequency of 2382.75 cm$^{-1}$, in good agreement with our calculated values of 2386--2390 cm$^{-1}$, suggesting that further experimental study of this system is worthwhile. The experimental reference for \ce{F2+} was taken from Cormack et al.\cite{cormack1996high} while Yang et al.\cite{yang2005combined}, consistent with Tentscher and Arey\cite{tentscher2012geometries}, suggest a value of 1091.5 cm$^{-1}$, further from our theoretical predictions. For \ce{HF+}, the large-basis results from Beran et al.\cite{beran2003approaching} and Tentscher and Arey\cite{tentscher2012geometries} predict an error that agree with our error prediction of 28 cm$^{-1}$. These calculated frequencies are closer to the value originally reported by Gewurtz et al.\cite{gewurtz1975electronic} and Hovde et al.\cite{hovde1989velocity} of 3090.5 cm$^{-1}$ adopted by Huber and Herzberg\cite{huber2018constants} than to the value of 3061.8 cm$^{-1}$ proposed by Yencha et al.\cite{yencha1999threshold}. Similarly, our predictions of the corrected vibrational frequency of \ce{HCl+} are blue-shifted by approximately 24 cm$^{-1}$ compared to the experimental reference value of 2673.69 cm$^{-1}$ from Irikura\cite{irikura2007experimental}. A recent joint experimental and theoretical study of this system by Patanen et al.\cite{patanen2014high} has suggested a computed value of 2702.6 cm$^{-1}$, in much better agreement with our results. For \ce{AlH+}, Huber and Herzberg\cite{huber2018constants} cite a 1934 study by Hoslt\cite{holst1934bandenspektrum} to approximate the vibrational frequency, though a more precise value is desired for the point of comparison to our computed frequencies to better assess the error between the different reference methods.    

\subsection{Pruned subset}
\begin{figure}
\centering
\caption{Box plots (overall, left, and enhanced, right) of the errors in corrected vibrational frequencies(in cm$^{-1}$) are presented for the pruned subset of species. Red lines mark the median errors, boxes bound the central 50\% of the data, whiskers enclose all data points within 1.5 times the inter-quartile range of the box edges, and points denote outlying data. }
\label{fig:boxplot_cropped}
\includegraphics[width=\linewidth]{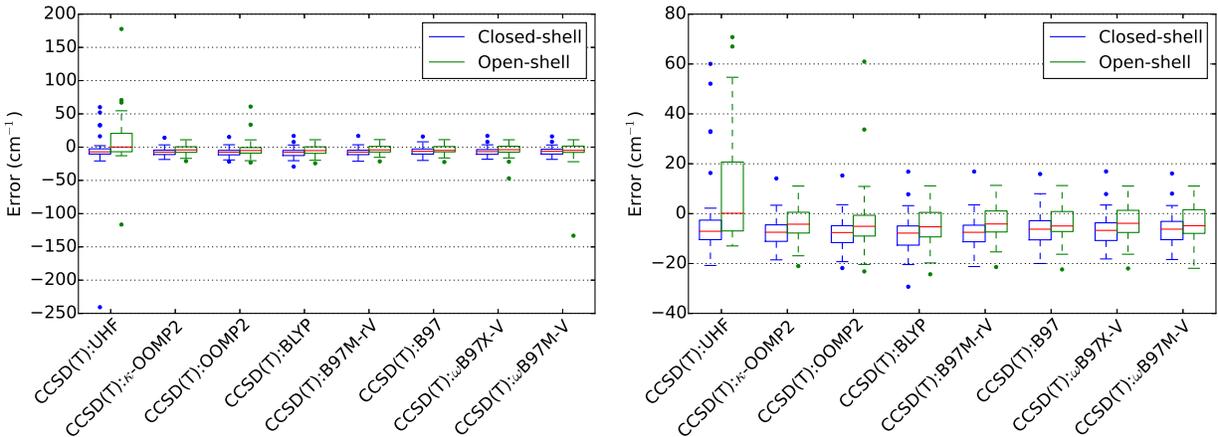}
\end{figure}

In order to draw more meaningful conclusions about the performance of CCSD(T) with $\kappa$-OOMP2 orbitals, we consider a subset of data points where the MR species (\ce{C2} and \ce{B2}) and the species with ambiguous experimental values (Tab. \ref{tab:questionable}) discussed above are excluded, leaving 35 closed-shell species (28 neutrals, 6 cations, 1 anion) and 46 open-shell species (36 neutrals, 9 cations, 1 anion; 35 doublets, 11 triplets).  Tab. \ref{tab:cropped} presents the RMSDs, MSDs, MINs, and MAXs for the closed-shell species, open-shell species, and the overall pruned set. These data are presented graphically in Fig. \ref{fig:boxplot_cropped}. For the pruned data set CCSD(T):B97 and CCSD(T):$\kappa$-OOMP2 are seen to yield the best performance with RMSDs of 8.48 cm$^{-1}$ and 8.50 cm$^{-1}$, respectively. The performances of CCSD(T):$\omega$B97X-V and CCSD(T):$\omega$B97M-V are hindered by \ce{C2-}; excluding this point brings the RMSDs for these methods to the same level as the other DFT-based methods.     

\begin{table}
\caption{Root mean square deviations, mean signed deviations, most negative deviations, and most positive deviations (all in cm$^{-1}$) for the cropped subset of species are summarized for the CCSD(T) methods utilizing different molecular orbitals.} \label{tab:cropped}
\tiny
\begin{tabular}{L{0.6cm}R{1.4cm}R{1.5cm}R{1.4cm}R{1.4cm}R{1.4cm}R{1.4cm}R{1.4cm}R{1.5cm}}
\hline
Closed-shell\newline pruned & \multicolumn{1}{C{1.4cm}}{$\Delta$(CCSD(T):\newline UHF)} & \multicolumn{1}{C{1.5cm}}{$\Delta$(CCSD(T):\newline $\kappa$-OOMP2)} & \multicolumn{1}{C{1.4cm}}{$\Delta$(CCSD(T):\newline OOMP2)} & \multicolumn{1}{C{1.4cm}}{$\Delta$(CCSD(T):\newline BLYP)} & \multicolumn{1}{C{1.4cm}}{$\Delta$(CCSD(T):\newline B97M-rV)} & \multicolumn{1}{C{1.4cm}}{$\Delta$(CCSD(T):\newline B97)} & \multicolumn{1}{C{1.4cm}}{$\Delta$(CCSD(T):\newline$\omega$B97X-V)} & \multicolumn{1}{C{1.5cm}}{$\Delta$(CCSD(T):\newline$\omega$B97M-V)} \\
\hline
RMSD & 44.43   & 9.41   & 10.18  & 11.56  & 10.05  & 9.45   & 9.50   & 8.97   \\
MSD  & -7.74   & -6.96  & -7.73  & -8.10  & -7.32  & -6.41  & -6.41  & -5.93  \\
MIN  & -240.71 & -18.53 & -21.83 & -29.30 & -21.21 & -20.00 & -18.16 & -18.40 \\
MAX  & 60.05   & 14.11  & 15.27  & 16.82  & 16.86  & 15.88  & 16.93  & 16.10  \\
\hline
Open-shell\newline pruned & \multicolumn{1}{C{1.4cm}}{$\Delta$(CCSD(T):\newline UHF)} & \multicolumn{1}{C{1.5cm}}{$\Delta$(CCSD(T):\newline $\kappa$-OOMP2)} & \multicolumn{1}{C{1.4cm}}{$\Delta$(CCSD(T):\newline OOMP2)} & \multicolumn{1}{C{1.4cm}}{$\Delta$(CCSD(T):\newline BLYP)} & \multicolumn{1}{C{1.4cm}}{$\Delta$(CCSD(T):\newline B97M-rV)} & \multicolumn{1}{C{1.4cm}}{$\Delta$(CCSD(T):\newline B97)} & \multicolumn{1}{C{1.4cm}}{$\Delta$(CCSD(T):\newline$\omega$B97X-V)} & \multicolumn{1}{C{1.5cm}}{$\Delta$(CCSD(T):\newline$\omega$B97M-V)} \\
\hline
RMSD & 38.33   & 7.75   & 13.45  & 8.80   & 7.61   & 7.67   & 10.26  & 21.11   \\
MSD  & 9.27    & -3.48  & -2.83  & -4.61  & -3.32  & -3.35  & -3.97  & -6.34   \\
MIN  & -116.46 & -21.02 & -23.17 & -24.32 & -21.41 & -22.38 & -46.93 & -133.20 \\
MAX  & 177.65  & 11.06  & 60.95  & 11.14  & 11.32  & 11.25  & 11.06  & 11.07   \\
\hline
Total\newline pruned  & \multicolumn{1}{C{1.4cm}}{$\Delta$(CCSD(T):\newline UHF)} & \multicolumn{1}{C{1.5cm}}{$\Delta$(CCSD(T):\newline $\kappa$-OOMP2)} & \multicolumn{1}{C{1.4cm}}{$\Delta$(CCSD(T):\newline OOMP2)} & \multicolumn{1}{C{1.4cm}}{$\Delta$(CCSD(T):\newline BLYP)} & \multicolumn{1}{C{1.4cm}}{$\Delta$(CCSD(T):\newline B97M-rV)} & \multicolumn{1}{C{1.4cm}}{$\Delta$(CCSD(T):\newline B97)} & \multicolumn{1}{C{1.4cm}}{$\Delta$(CCSD(T):\newline$\omega$B97X-V)} & \multicolumn{1}{C{1.5cm}}{$\Delta$(CCSD(T):\newline$\omega$B97M-V)} \\
\hline
RMSD & 41.07   & 8.50   & 12.15  & 10.08  & 8.75   & 8.48   & 9.94   & 16.96   \\
MSD  & 1.92    & -4.98  & -4.95  & -6.11  & -5.04  & -4.67  & -5.03  & -6.17   \\
MIN  & -240.71 & -21.02 & -23.17 & -29.30 & -21.41 & -22.38 & -46.93 & -133.20 \\
MAX  & 177.65  & 14.11  & 60.95  & 16.82  & 16.86  & 15.88  & 16.93  & 16.10   \\
\hline
\end{tabular}
\end{table}

Comparing Fig. \ref{fig:boxplot_all} and Fig. \ref{fig:boxplot_cropped}, we see that the pruning procedure removed many of the outlier cases that were shared between all or nearly all of the methods surveyed. For CCSD(T):UHF, the remaining outlying points are the spin-contaminated points from the closed-shell subset (\ce{F2}, \ce{PN}, \ce{BeS}, \ce{NS+}, and \ce{P2}), and \ce{HCl}, while the remaining open-shell outliers are \ce{NO}, \ce{N2+}, \ce{CO+}, and \ce{C2-}. The range covered by the CCSD(T):UHF whiskers is larger by a factor of two than nearly all the other methods tested. For CCSD(T):$\kappa$-OOMP2, the pruning procedure leaves only one closed-shell and one open-shell outlier, \ce{HCl} and \ce{BO}, respectively. For CCSD(T):OOMP2, the remaining outliers in the closed-shell subset are \ce{HCl} and \ce{BeO} while the open-shell set has outliers \ce{P2+}, which has been previously discussed, \ce{Si2}, which breaks spatial symmetry at the OOMP2 level, and \ce{BO}. CCSD(T):BLYP has closed-shell outliers in \ce{NaLi}, \ce{NaH}, and \ce{HCl} and an open-shell outlier in \ce{BO}.  CCSD(T):B97M-rV and CCSD(T):B97 both have \ce{HCl} as an open-shell outlier and \ce{BO} as a closed-shell outlier. CCSD(T):$\omega$B97X-V has closed-shell outliers in \ce{HCl} and \ce{NaH} and open-shell outliers in \ce{C2-} and \ce{BO}. CCSD(T):$\omega$B97M-V also has \ce{HCl} and \ce{NaH} as closed-shell outliers and \ce{C2-} as an open-shell outlier. The non-HF methods are seen to have their mean values red-shifted compared to experiments by 5-6 cm$^{-1}$. 

\section{Conclusions}
We have evaluated the performance of CCSD(T) with different MO references to predict the vibrational frequencies of both closed-shell and open-shell diatomic molecules and ions. The at times problematic use of a UHF reference was compared against the use of $\kappa$-OOMP2, OOMP2, BLYP, B97M-rV, B97, $\omega$B97X-V, and $\omega$B97M-V molecular orbital references. Overall, CCSD(T):B97, CCSD(T):$\kappa$-OOMP2, and CCSD(T):B97M-rV yield RMSDs on the pruned overall data set of 8.48 cm$^{-1}$, 8.50 cm$^{-1}$, and 8.75 cm$^{-1}$, respectively, reducing the RMSD for CCSD(T):UHF by nearly a factor of 5. For the pruned closed- and open-shell subsets the associated RMSDs are 9.45 cm$^{-1}$ and 7.67 cm$^{-1}$, respectively, 9.41 cm$^{-1}$ and 7.75 cm$^{-1}$, respectively, and 10.05 and 7.61, respectively, for CCSD(T):B97, CCSD(T):$\kappa$-OOMP2, and CCSD(T):B97M-rV. The slightly degraded performance of the other non-HF MO references are skewed by one or two data points and otherwise rank competitively with the B97, $\kappa$-OOMP2, and B97M-rV. These outlying data points are seen to arise from spatial or spin-symmetry breaking or erroneous symmetry restoration. The effect of regularization in $\kappa$-OOMP2 is seen to prevent the symmetry issues seen in OOMP2. 

A major practical limitation of these very promising non-HF CCSD(T) methods is the present lack of implemented analytic gradients, relegating the current application of these to systems with only a few atoms. One may develop an approach like that of Taube and Bartlett's FNO-CCSD(T)\cite{taube2008frozen} for $\kappa$-OOMP2 and/or DFT orbitals. The success of the non-HF CCSD(T) methods in treating closed-shell and open-shell systems with the same accuracy speaks to the usefulness of such an implementation. Such approaches do not affect the overall asymptotic scaling and would extend the "black box" utility of CCSD(T) currently seen for closed-shell systems. Furthermore, the use of $\kappa$-OOMP2 as the generator of MOs also provides a diagnostic tool in the mean-field $\langle S^{2} \rangle$ of the multireference character of the target system, informing the expected accuracy of the subsequent CCSD(T) results. 

In order to assess errors relating to the computational treatment presented in this work we consider errors in the basis set and errors related to the approximate treatment of triple excitations. Peterson and Dunning\cite{peterson2002accurate} demonstrated a blue shift of computed harmonic frequencies for row 3 diatomic species of 2-15 cm$^{-1}$ when going from the cc-pwCVTZ to cc-pwCV5Z. Provided this trend holds with the aug-cc-pwCV$n$Z basis sets the use of the aug-cc-pwCVQZ basis should help to correct the systematic red-shift seen for the non-HF CCSD(T) methods. For exact treatment of triples Tentscher and Arey\cite{tentscher2012geometries} found CCSDT tends to red-shift in computed frequencies compared to CCSD(T). This effect is exacerbated for species where CCSD(T) with UHF orbitals is particularly poor (\ce{N2+}, \ce{CN}) while for other other species the shift was on the order of 5 cm$^{1}$. We also note that CCSDT, free of the perturbative nature of CCSD(T), should be less sensitive to the choice of reference orbitals.

Further extension of this study approach to cases where DFT traditionally struggles would be of significant interest. In particular, extension to transition metals species and systems with significant charge separation is highly desirable to discriminate if CCSD(T):DFT is able to remedy the traditional failures of DFT. Additionally, the exponential regularization schemes of Lee and Head-Gordon\cite{lee2018regularized} could be extended to the perturbative triples calculation in CCSD(T) as an attempt to handle cases of nonvariational failure in traditional CCSD(T).



\begin{acknowledgement}
This work was supported by the Director, Office of Science, Office of Basic Energy Sciences, of the U.S. Department of Energy under Contract No. DE-AC02-05CH11231. L. W. B. thanks the NSF for a NSF Graduate Research Fellowship DGE-1106400. 

\end{acknowledgement}

\begin{suppinfo}
Tables of equilibrium bond lengths, comments on the data set, mean-field $\langle S^{2}\rangle$ values, and CCSD data are presented in the Supporting Information.  
\end{suppinfo}

\bibliography{ccsdt_frequencies}

\end{document}